\documentclass{Interspeech2024}
\usepackage{multirow}
\usepackage{tabularx}
\interspeechcameraready
\title{Pitch-Aware RNN-T for Mandarin Chinese Mispronunciation Detection and Diagnosis}

\name[]{Xintong}{Wang}
\name[]{Mingqian}{Shi}
\name[]{Ye}{Wang}

\address{
  School of Computing, National University of Singapore, Singapore}
\email{\{wangxt,m-shi,wangye\}@comp.nus.edu.sg}


\begin{document}

\maketitle

\begin{abstract}
Mispronunciation Detection and Diagnosis (MDD) systems, leveraging Automatic Speech Recognition (ASR), face two main challenges in Mandarin Chinese: 1) The two-stage models create an information gap between the phoneme or tone classification stage and the MDD stage. 2) The scarcity of Mandarin MDD datasets limits model training. In this paper, we introduce a stateless RNN-T model for Mandarin MDD, utilizing HuBERT features with pitch embedding through a Pitch Fusion Block. Our model, trained solely on native speaker data, shows a 3\% improvement in Phone Error Rate and a 7\% increase in False Acceptance Rate over the state-of-the-art baseline in non-native scenarios.
\end{abstract}

\section{Introduction}
While learning a second language (L2), learners may produce mispronunciations due to various factors, such as the influence of their first language (L1)~\cite{zhang2023phonetic}. Mispronunciations may manifest at the segmental level, involving phonemes, or at the supra-segmental level, which encompasses aspects such as prosody, fluency, and intonation. Mispronunciation Detection and Diagnosis (MDD) systems are utilized to identify these pronunciation errors and provide automatic feedback~\cite{li2016mispronunciation}. In this work, we focus on the pronunciation errors of Mandarin Chinese learners, taking into account the unique challenges posed by its tonal nature.

Traditional methods for pronunciation assessment involve calculating variations of log-posterior probability to derive pronunciation scores, such as the Goodness of Pronunciation (GOP)~\cite{witt2000phone}, scaling log-posterior probabilities~\cite{zhang2008automatic}, and evaluating the log-likelihood ratio~\cite{franco1999automatic}. Although these approaches are relatively intuitive, they exhibit limitations in accuracy. This deficiency stems from the uniform scoring of all speech without accommodating the distinctive acoustic-phonetic characteristics of individual utterances~\cite{truong2004automatic}. In efforts to enhance performance, researchers have developed classifiers tailored to specific pronunciation errors. Additionally, Harrison et al.~\cite{harrison09_slate} introduced the Extended Recognition Network (ERN), incorporating 51 context-sensitive phonological rules represented as finite state transducers. This ERN significantly improves the accuracy of pronunciation assessment. Nonetheless, this method faces challenges in fully addressing the diverse range of pronunciation error types.

Recently, Leung et al.~\cite{leung2019cnn} introduced a CNN-RNN-CTC model that leverages end-to-end Automatic Speech Recognition (ASR) for MDD and demonstrated superior performance over ERN-based models without the need for phonemic or graphemic information, or forced alignment between different linguistic units. Subsequently, Zhang et al.~\cite{zhang2023phonetic} adopted an autoregressive model, the Recurrent Neural Network Transducer (RNN-T)~\cite{graves2012sequence}, for MDD. This approach aims to capture the temporal dependence of mispronunciation patterns, showing better performance than Connectionist Temporal Classification (CTC)-based methods. Xu et al.~\cite{xu2021explore} also found that applying CTC loss directly, without canonical phoneme information, yielded worse results, likely due to the lack of textual context. However, with insufficient mispronunciation patterns in the training data, models tend to predict phonemes following canonical linguistic rules. Ghodsi et al.~\cite{ghodsi2020rnn} proposed a stateless RNN-T model that replaces the recurrent neural network with simple non-autoregressive layers while maintaining comparable accuracy in ASR tasks. The reduction in parameters in stateless RNN-T models has also accelerated training speed.

Data sparsity, highlighting the scarcity of annotated non-native speech data, is a critical issue in Mandarin Chinese MDD. Established datasets such as Speechocean762~\cite{zhang2021speechocean762} and L2-ARCTIC~\cite{zhao2018l2} have significantly advanced MDD research for L2 English. However, for L2 Mandarin Chinese, the lack of sufficient data impedes the development of robust ASR-based MDD systems. To our knowledge, the only publicly available L2 Mandarin Chinese dataset for training is the relatively small LATIC dataset~\cite{zhao2021latic}. Most studies in Mandarin Chinese MDD, including those by Chen et al.~\cite{chen2004automatic}, Hu et al.~\cite{hu2015improved}, Shen et al.~\cite{shen2022self}, and Guo et al.~\cite{guo2023multi}, have relied heavily on private datasets. In scenarios of limited data availability, Self-Supervised Learning (SSL) models pre-trained on large unlabeled datasets demonstrate significant potential in MDD, as evidenced by Liu et al.~\cite{liu2023zero}. Similarly, Shen et al.~\cite{shen2022self} utilized SSL models to train an ASR-based model for MDD in Mandarin Chinese. However, this method did not explicitly extract pitch information from speech, which has been shown to enhance the performance of MDD models in tonal languages~\cite{huu23_interspeech}.

In this paper, we propose an approach that involves the fine-tuning of HuBERT~\cite{hsu2021hubert} with a stateless RNN-T for Mandarin Chinese MDD. Simultaneously, F0 is extracted from the waveform to generate pitch embedding, which is then fed into a pitch encoder to obtain high-dimensional pitch features. A Pitch Fusion Block is utilized by the model to combine HuBERT features with pitch features, aiming to improve MDD performance. Our proposed model was trained on AISHELL-1~\cite{bu2017aishell} and evaluated on the LATIC~\cite{zhao2021latic} dataset. The results demonstrate that our model achieved comparable performance to other models and showed a 3\% relative improvement in the Phone Error Rate and a 7\% increase in the False Acceptance Rate compared to a state-of-the-art baseline.

\section{Method}
\begin{figure}[t]
  \centering
  \includegraphics[width=\linewidth, trim=2.5cm 4cm 10cm 1cm, clip]{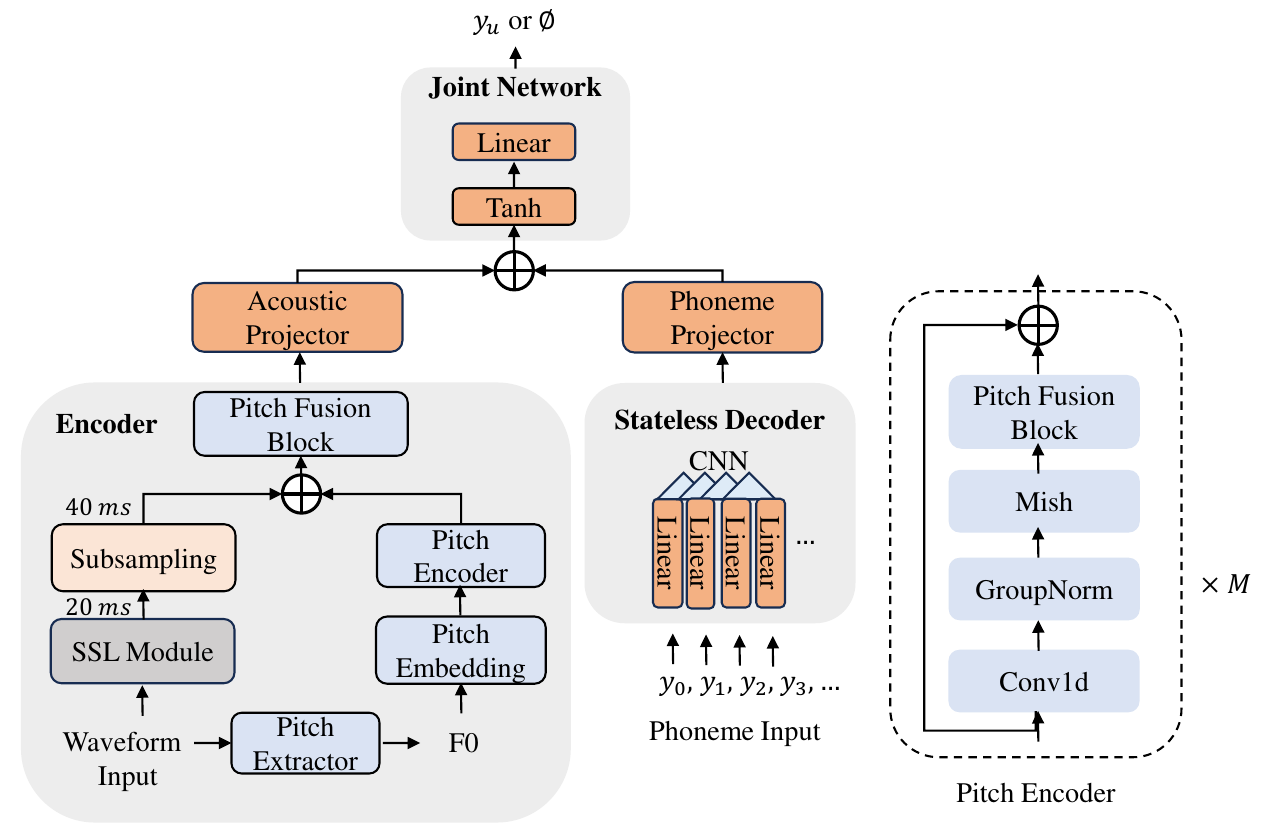}
  \caption{The Proposed Tonal Phoneme MDD Framework.}
  \label{fig:model}
\end{figure}

Our model, as shown in Figure~\ref{fig:model}, follows a stateless RNN-T architecture, with an Encoder, a Stateless Decoder, and a Joint Network.
The Encoder comprises an SSL module based on HuBERT, a Subsampling module, a Pitch Extractor, a Pitch Embedding, a Pitch Encoder, and a Pitch Fusion Block.
The Stateless Decoder and Joint Network used in our model follow the structure outlined in~\cite{yao2023zipformer}.

\subsection{Stateless RNN-T Overview}
The original RNN-T comprises three key components: an Encoder, a Prediction Network (also referred to as a Decoder), and a Joint Network. Given a length $L$ input acoustic feature sequence, such as MFCCs or Fbanks, denoted as $\boldsymbol{f} = (f_1, \ldots, f_L)$, and a phoneme sequence $\boldsymbol{y}$ of length $U+1$, $\boldsymbol{y} = (y_0, \ldots, y_U)$, over the phoneme set $\mathcal{P}$. The Encoder maps $\boldsymbol{f}$ into a high-dimensional acoustic representation. The Decoder is an autoregressive model that encodes $y_{0...u-1}$ ($u \in \{1, 2, 3, \ldots, U-1\}$) into a high-dimensional phoneme representation. The encoder output and the decoder output are then projected to the same size. Subsequently, the Joint Network combines them to jointly predict $y_u$ or $\varnothing$, where the blank token $\varnothing$ signifies nothing from $\mathcal{P}$ outputted at the current token position. $y_0$ with $\varnothing$ represents the start of the sentence. The RNN-T loss $\mathcal{L}_{RNN-T}$ is defined as:

$$
\mathcal{L}_{RNN-T} = - P (\boldsymbol{y} | \boldsymbol{f}) = - \sum_{\boldsymbol{a} \in \boldsymbol{M^{-1}} (\boldsymbol{y})} P (\boldsymbol{a} | \boldsymbol{f}),
$$
where $\boldsymbol{a}$ refers to an alignment between $\boldsymbol{y}$ and $\boldsymbol{f}$. $\boldsymbol{a}$ is a frame-level phoneme sequence with its length as $L$. The various locations of the blank tokens refer to different alignments. $\boldsymbol{M}$ is a function that removes the blank tokens from $\boldsymbol{a}$. The model is optimized by maximizing the summation of probabilities of all alignments.

\subsection{Phonetic Representation}
For the MDD task, we use tonal phonemes to assess pronunciation with greater granularity. We use an open-source lexicon~\footnote{https://www.mdbg.net/chinese/dictionary?page=cc-cedict}, which encompasses most of the commonly used Chinese words and characters. This lexicon is also adopted in the AISHELL-1 dataset~\cite{bu2017aishell}. 
Pronunciations are represented using the initial-final-tone system, where syllables are broken down into their initial consonant sounds, final vowel sounds, and tones.
This phoneme set includes five tones: Tone 1 (high), Tone 2 (rising), Tone 3 (falling then rising), Tone 4 (high then falling), and Tone 5 (neutral or toneless)~\cite{tong2016context}. Notably, we regard zero-initial syllables as single tonal finals in this work.

\subsection{SSL module}
We utilize HuBERT as the SSL module in our model. This module is employed for encoding speech from the waveform, aiming to provide enhanced representations. During implementation, the SSL module has a down-sampling factor of 320, equivalent to a 20ms hop size for audio sampled at 16kHz~\cite{hsu2021hubert}. In this work, we adopt a Subsampling module at the top layer of HuBERT to achieve a 40ms hop size for the output feature. This involves concatenating each two successive frames and then applying a linear layer with a Tanh activation function.

\subsection{Pitch Extractor}
To provide pitch information, the fundamental frequency (F0) is estimated with DIO~\cite{morise2009fast} in WORLD~\cite{morise2016world}. 
We analyze the distribution of F0 values across all training frames after applying speed perturbation at factors of 0.9, 1.0, and 1.1 using Lhotse~\cite{zelasko2021lhotse}. 
We observe that most of the F0 values fall in the range of 100 - 600 Hz, with 100 - 200 Hz being the most common, and rarely exceeding 600 Hz, resulting in an unbalanced distribution. Therefore, we further apply Mel-scaling to the extracted F0, along with min-max normalization and discretization to obtain a coarse F0 with bins of size 256. We introduce an embedding layer (Pitch Embedding in Figure~\ref{fig:model}) to map the F0 or variants of F0 into a higher-dimensional representation before feeding them into the Pitch Encoder. F0 is extracted with various hop sizes, including 10 ms, 20 ms, and 40 ms. Experiments were conducted to explore the impact of pitch extraction on the model's performance, as detailed in Section~\ref{sec:pitch_extraction}.

\subsection{Pitch Fusion Block}
\label{sec:pitch_fusion_block}
\begin{figure}[t]
  \centering
  \includegraphics[width=0.8\linewidth, trim=6cm 4cm 11cm 2cm, clip]{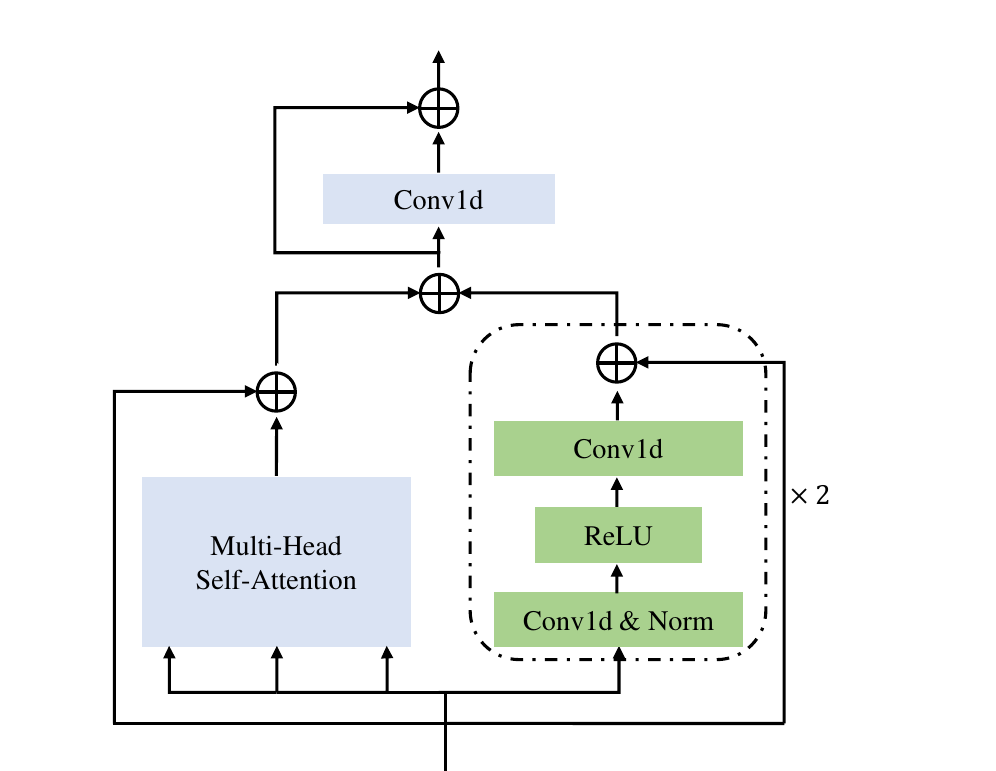}
  \caption{The architecture of the Pitch Fusion Block. The Multi-Head Self-Attention is designed to capture global pitch features, while the residual convolution blocks (delineated by dotted lines and colored in green) aim to capture local pitch features.}
  \label{fig:pitchfusionblock}
\end{figure}

The Pitch Fusion Block is used to fuse the extracted pitch features and HuBERT features.
In Mandarin Chinese, the tone of an individual character can be determined by its short-term F0 contour. Meanwhile, this tonal identity is also subject to modification by the tones of preceding characters, a phenomenon known as tone sandhi.
Hence, the Pitch Fusion Block is used to synergize the modeling of long-range global features with the detailed local feature patterns observed in the F0 contour. As shown in Figure~\ref{fig:pitchfusionblock}, we implemented the Pitch Fusion Block with Multi-Head Self-Attention to capture global features and residual convolution blocks for extracting local features. Subsequently, we sum the global and local features and normalize the resultant output. This is similar to the ConvFFT block presented in~\cite{wang2022xiaoicesing, wang2023crosssinger}.

\subsection{Pitch Encoder}
We implement the Pitch Encoder with a 1-D convolutional layer with group normalization~\cite{wu2018group} and the Mish~\cite{misra2019mish} activation function. Subsequently, the Pitch Encoder employs a Pitch Fusion Block (Section~\ref{sec:pitch_fusion_block}) implemented with Multi-Head Self-Attention and residual convolution blocks. In the experiments, $M$ Pitch Encoders are concatenated to accommodate different pitch extraction hop sizes (see Section~\ref{sec:experimental_settings}).

\section{Experiments}
\subsection{Datasets}
We train the MDD models using the AISHELL-1 corpus and evaluate their performance with the LATIC dataset. LATIC, annotated by human experts, is a non-native Mandarin Chinese speech dataset utilized to assess the efficacy of L2 MDD methods.
The LATIC dataset comprises recordings from four speakers, each with a different L1 language: Russian, Korean, French, and Arabic. Dataset statistics are summarized in Table~\ref{tab:dataset}.

\begin{table}[th]
      \caption{Datasets Summary: The number of speakers, duration, utterances (Utt.), and L1 of speakers in AISHELL-1 and LATIC Datasets.}
  \label{tab:dataset}
  \centering
  \begin{tabularx}{0.48\textwidth}{lcccc}
    \toprule
    \multirow{2}{*}{} & \multicolumn{3}{c}{\textbf{AISHELL-1}} &        
                                \multicolumn{1}{c}{\textbf{LATIC}} \\
                      & Train & Dev & Test & Test \\
    \midrule
    Speakers          & 340   &  40 & 20  &   4        \\
    Hours             & 150   & 10  & 5   &  4           \\
    Utt.        & 120098 & 14326  &  7176 & 2579                \\
    L1       & \multicolumn{3}{c}{Mandarin Chinese}        & \begin{tabular}[c]{c}Russian, Korean, \\ French, and Arabic \end{tabular} \\
    \bottomrule
  \end{tabularx}
  
\end{table}

\begin{table*}[h]
  \caption{Overall Performance Comparison of PER and MDD Metrics. ``Hop Size'' indicates the hop size during pitch extraction. ``Pitch Encoder'' indicates whether the Pitch Encoder of the model is implemented with a Pitch Fusion Block (PFB). ``PE'' indicates models implemented with Pitch Embedding in the encoder.}
  \label{tab:results}
  \centering
  \begin{tabularx}{\textwidth}{Xcccccccl}
    \toprule
    \textbf{Model} & \textbf{Hop Size} & \textbf{Pitch Encoder} & \textbf{PER} $\downarrow$ & \textbf{FRR} $\downarrow$ & \textbf{FAR} $\downarrow$ & \textbf{Pre.} $\uparrow$ & \textbf{Rec.} $\uparrow$ & \textbf{F1-score} $\uparrow$\\
    \midrule
    \textbf{Baseline} \\
    wav2vec2.0-CTC~\cite{xu2021explore} & - & -  & 27.55 & 0.266 & 0.083 & 0.109 & 0.917 & 0.195 \\
    \midrule
    \textbf{Stateless RNN-T} \\
    Pre-trained HuBERT  & - & - & 28.65 & 0.274 & \textbf{0.063} & 0.108 & \textbf{0.937}& 0.194 \\
    Fine-tuned HuBERT & - & - & 27.22 & 0.261 & 0.074 & 0.109 & 0.926 & 0.196 \\
        - Raw F0 (linear)  & 10 ms & w/o PFB & 31.74 & 0.303 & 0.075 & 0.082 & 0.925 & 0.150 \\ 
        - Raw F0 w/ PE (linear)  & 10 ms & w/o PFB & 27.32 & 0.263 & 0.087 & 0.111 & 0.913 & 0.197 \\
        - Mel-scaled F0 w/ PE (linear) & 10 ms & w/o PFB & 27.46 & 0.264 & 0.087 & \textbf{0.113} & 0.913 & 0.200 \\
        - Coarse F0 w/ PE (linear) & 10 ms & w/o PFB & 27.46 & 0.264 & 0.088 & 0.109 & 0.912 & 0.195 \\
        - Raw F0 w/ PE (linear)  & 10 ms & w/ PFB & 27.28 & 0.263 & 0.085 & 0.112 & 0.915 & 0.200 \\
        - Raw F0 w/ PE (linear)  & 20 ms & w/ PFB & 27.27 & 0.261 & 0.080 & 0.113 & 0.920 & \textbf{0.201} \\
        - Raw F0 w/ PE (linear)  & 40 ms & w/ PFB & 27.22 & 0.262 & 0.088 & 0.112 & 0.912 & 0.200 \\
        \quad - Pitch Fusion Block (global)  & 40 ms & w/ PFB & 27.25 & 0.263 & 0.085 & 0.112 & 0.915 & 0.200 \\
        \quad - Pitch Fusion Block & 40 ms & w/ PFB & \textbf{26.69} & \textbf{0.257} & 0.077 & 0.111 & 0.922 & 0.198 \\
        Fine-tuned wav2vec2.0 &   &   &    &   &   &   &  &  \\
        - Raw F0 w/ PE &   &   &    &   &   &   &  &  \\
        \quad \quad - Pitch Fusion Block & 40 ms & w/ PFB &  27.54 & 0.266 & 0.077 & 0.103 & 0.923 & 0.185 \\
    \bottomrule
  \end{tabularx}
\end{table*}

\subsection{Experimental Settings}
\label{sec:experimental_settings}
We employ pre-trained chinese-wav2vec2-base and chinese-hubert-base by TencentGameMate\footnote{https://github.com/TencentGameMate/chinese\_speech\_pretrain} for the SSL module. The subsampling output dimension is set to 1024. The input size of the pitch embedding layer is configured to be 1600, a parameter determined by conducting a statistical analysis on the maximum F0 observed within the training dataset. The Pitch Embedding is designed with an embedding size of 512. 
Our Pitch Encoder is implemented with three configurations. 
We use $M$ to denote the number of Pitch Encoders concatenated during implementation. Specifically, when the hop size for F0 extraction is set to 10 ms, $M$ is configured as 2, with the stride for the 1-D convolutional layers being 2. For a hop size of 20 ms, $M$ is reduced to 1, maintaining a stride of 2. For a hop size of 40 ms, $M$ remains 1, but the stride is adjusted to 1. These configurations ensure that the output size of the Pitch Encoder matches that of the HuBERT features.
The hyperparameters of the Pitch Fusion Block used in this work are outlined in~\cite{wang2022xiaoicesing}. We set the embedding dimension to 1024 and use 4 attention heads. The vocabulary size in our work is 215, including 214 tonal phonemes tokens and a blank token. The acoustic projector maps the 1024-dimensional acoustic feature into 512, while the phoneme projector maps the phoneme embedding into 512. The architectural details of the stateless decoder and the Joint network are delineated in~\cite{yao2023zipformer}.

We fine-tune wav2vec2.0-CTC~\cite{xu2021explore} as the baseline model and compare it with the proposed stateless RNN-T-based models using the k2 framework\footnote{https://github.com/k2-fsa/icefall} and Fairseq\footnote{https://github.com/facebookresearch/fairseq}. The SSL modules were frozen for the initial 10,000 training steps and subsequently commenced fine-tuning after this initial phase. By default, we set the max-duration per GPU to 100s in k2 and fine-tuned the SSL modules for 20 epochs. All models are trained on 24GB NVIDIA RTX A5000 GPU. We fine-tuned SSL modules using the same optimization strategy as mentioned in~\cite{yao2023zipformer}. During decoding, greedy search is employed.

\subsection{Metrics and Overall Experimental Results}
Following previous works~\cite{qian2010discriminative,yan2020end}, we employ several evaluation metrics to assess the performance of the MDD model, including False Rejection Rate (FRR; FR/(FR + TA)), False Acceptance Rate (FAR; FA/(FA + TR)), Recall (RE; TR/(FA + TR)), Precision (PR; TR/(FR + TR)), and F1-score (2*(RE * PR)/(RE + PR)). The True Rejection (TR) represents the number of phonemes labeled as mispronunciations and detected as incorrect. False Rejection (FR) is the number of phonemes annotated as correct pronunciation and identified as incorrect. False Acceptance (FA) is the number of phonemes that are mispronounced but misclassified as correct. True Acceptance (TA) is the number of correct pronounced phonemes classified as correct. Additionally, we compute the Phoneme Error Rate (PER) to evaluate the performance of the phoneme recognition model. 

Experimental results are summarized in Table~\ref{tab:results}. It is observed that L1 fine-tuned HuBERT achieves improvements in PER, FRR, and F1-score among the HuBERT model initialized by the pre-trained parameters and the baseline. 

To assess the efficacy of wav2vec2.0~\cite{baevski2020wav2vec} and HuBERT in the MDD task, we fine-tuned wav2vec2.0 under the same configuration as HuBERT in our best model. The experimental results reveal that the proposed stateless RNN-T with HuBERT achieves a notable improvement in the PER, and FRR, Precision, and F1-score, outperforming the wav2vec2.0 based model with the same stateless RNN-T architecture.

\subsection{Effectiveness of Different Pitch Extraction Methods}
\label{sec:pitch_extraction}
We conduct a series of experiments to evaluate the efficacy of pitch extraction methods (Table~\ref{tab:results}). 
Using the same pitch extraction hop size (10 ms) and Pitch Encoder (w/o PFB), the raw F0 with pitch embedding (Raw F0 w/ PE as shown in Table~\ref{tab:results}) achieves a lower PER and FRR, outperforming models that utilize mel-scaled F0 and coarse F0. Furthermore, compared to the model that uses raw F0 without Pitch Embedding, the model with raw F0 and Pitch Embedding achieves a 16\% reduction in PER, a 35\% improvement in Precision, and a 31.3\% improvement in F1-score. This demonstrates the effectiveness of using a high-dimensional representation of raw F0 over raw F0 alone for Mandarin Chinese MDD.

The results for various hop sizes, specifically at 10 ms, 20 ms, and 40 ms, are presented in Table~\ref{tab:results}. The results indicate that the model with a 40 ms hop size for F0 as input achieved the lowest PER.

\subsection{Effectiveness of Different Pitch Fusion Methods}
We compare three models with different pitch fusion methods fusing Pitch Encoder output with HuBERT features. The proposed model with complete Pitch Fusion Block is denoted as the ``Pitch Fusion Block'' in Table~\ref{tab:results}. 
For comparison, we replace the Pitch Fusion Blocks (in Figure~\ref{fig:model}) with a linear layer. These models are marked with ``linear'' in Table~\ref{tab:results}.
Furthermore, we remove the convolution residual blocks in the Pitch Fusion Block (delineated by dotted lines and colored in green in Figure~\ref{fig:pitchfusionblock}) to evaluate the effect of extracted local features on MDD performance. As the remaining Multi-Head Self-Attention extracts the global features, we mark this model with ``global'' in Table~\ref{tab:results}.
Results show that the proposed model, which incorporates pitch-aware methodologies, reduces the PER and FRR, and achieves higher precision and F1-score compared to the fine-tuned HuBERT without pitch input. 

We list models with high Recall (higher than 0.922) in Table~\ref{tab:count}, and provide more detailed metrics to analyze their performance.
True Rejection (TR) consists of two components: Correct Diagnosis (CD) and Diagnostic Errors (DE). 
CD represents the count of mispronunciations accurately identified by the model. DE refers to the instances where mispronunciations are correctly detected but incorrectly attributed to a different phoneme than the one actually produced by the L2 speaker. 
For instance, if the expected phoneme is `sh' and the L2 speaker pronounces it as `s', a model recognition of `s' would be classified under CD, indicating a correct diagnosis. Conversely, if the model incorrectly identifies the mispronounced phoneme as `c', this instance would be categorized as a DE, highlighting a correct detection but erroneous recognition.
We further adopt the Diagnostic Error Rate (DER; DE / CD + DE) proposed in~\cite{li2016mispronunciation} to measure the performance of our model. As demonstrated in Table~\ref{tab:count}, our model exhibits the lowest DER, signifying a reduced incidence of Diagnostic Errors in True Rejections when compared to competing models.

\begin{table}[th]
  \caption{Comparison of Diagnostic Error Rate (DER) in models with high recall.}
  \label{tab:count}
  \centering
  \begin{tabularx}{0.47\textwidth}{lcccc}
    \toprule
    \textbf{Model} & \textbf{Hop Size} & \textbf{Pitch} & \textbf{DER $\downarrow$} \\
    \textbf{} & \textbf{} & \textbf{Encoder} & \textbf{} \\
    \midrule
    Pre-trained HuBERT & - & - & 0.321 \\
    Fine-tuned HuBERT & - & - & 0.320 \\
    - Raw F0 (linear) & 10 ms & w/o PFB & 0.370 \\
    - Raw F0 w/ PE &   &   &  \\
    \quad - Pitch Fusion Block & 40 ms & w/ PFB & \textbf{0.318}                \\
    \bottomrule
  \end{tabularx}
\end{table}

\vspace{-10pt}
\section{Conclusion}
This paper introduces a pitch-aware Recurrent Neural Network Transducer specifically designed for Mandarin Chinese Mispronunciation Detection and Diagnosis. The proposed model employs a novel fusion methodology that integrates pitch embeddings with HuBERT features to achieve state-of-the-art performance. Additionally, this study investigates the impact of various hop sizes on F0 extraction, the use of Mel-scaled F0, and different pitch fusion mechanisms on model performance. We anticipate that the findings presented herein will serve as a catalyst for future research in the areas of tonal language Automatic Speech Recognition and Mispronunciation Detection and Diagnosis.

\section{Acknowledgements}
The authors would like to thank anonymous reviewers for their
valuable suggestions. This project is funded by a research grant MOE-MOESOL2021-0005 from the Ministry of Education in Singapore.

\bibliographystyle{IEEEtran}
\bibliography{mybib}

\end{document}